\documentclass{article}

\usepackage{arxiv}

\usepackage[utf8]{inputenc} % allow utf-8 input
\usepackage[T1]{fontenc}    % use 8-bit T1 fonts
\usepackage{hyperref}       % hyperlinks
\usepackage{url}            % simple URL typesetting
\usepackage{booktabs}       % professional-quality tables
\usepackage{amsfonts}       % blackboard math symbols
\usepackage{nicefrac}       % compact symbols for 1/2, etc.
\usepackage{microtype}      % microtypography
\usepackage{cleveref}       % smart cross-referencing
\usepackage{lipsum}         % Can be removed after putting your text content
\usepackage{graphicx}
\usepackage{natbib}
\usepackage{doi}

\title{Interactions between irregular wave fields and sea ice: A physical model for wave attenuation and ice breakup in an ice tank}

\date{}

\author{Giulio Passerotti \\
	    The University of Melbourne, Australia\\
	    \texttt{giuliopasserotti@gmail.com} \\
	    \And
    	Luke G.~Bennetts \\
    	University of Adelaide, Australia
    	\And
    	Franz von Bock und Polach \\ 
    	Hamburg University of Technology, Hamburg, Germany \\
    	Universität Hamburg, Hamburg, Germany
    	\And
    	Alberto Alberello \\
    	University of East Anglia, Norwich, UK
    	\And
    	Otto Puolakka \\
    	Aalto University, Helsinki, Finland
    	\And
    	Azam Dolatshah \thanks{Azam Dolatshah's current affiliation: BMT Commercial Australia Pty Ltd, Australia.} \\
    	The University of Melbourne, Australia \\
    	\And
    	Jaak Monbaliu \\
        KU Leuven, Leuven, Belgium
        \And
        Alessandro Toffoli \\
        The University of Melbourne, Australia \\
        \texttt{toffoli.alessandro@gmail.com}}

\renewcommand{\shorttitle}{\textit{arXiv} Template}

\hypersetup{
pdftitle={A template for the arxiv style},
pdfsubject={q-bio.NC, q-bio.QM},
pdfauthor={David S.~Hippocampus, Elias D.~Striatum},
pdfkeywords={First keyword, Second keyword, More},
}

\begin{document}
\maketitle

\begin{abstract}
    Irregular, unidirectional surface water waves incident on model ice in an ice tank are used as a physical model of ocean surface wave interactions with sea ice. Results are given for an experiment consisting of three tests, starting with a continuous ice cover and in which the incident wave steepness increases between tests. The incident waves range from causing no breakup of the ice cover to breakup of the full length of ice cover. Temporal evolution of the ice edge, breaking front and mean floe sizes are reported. Floe size distributions in the different tests are analysed. The evolution of the wave spectrum with distance into the ice-covered water is analysed in terms of {changes of} energy content, mean wave period and spectral bandwidth {relative to their incident counterparts}, and pronounced differences are found between the tests. Further, an empirical attenuation coefficient is derived from the measurements and shown to have a power-law dependence on frequency comparable to that found in field measurements. Links between wave properties and ice breakup are discussed.
\end{abstract}

% \keywords{First keyword \and Second keyword \and More}

%%%%%%%%%%%%%%%%%%
%% INTRODUCTION %%
%%%%%%%%%%%%%%%%%%

\section{Introduction} \label{sec:intro}

Over the past decade, interactions between surface gravity waves and sea ice have become more broadly recognised as important for wave and sea ice processes \citep{squire_past_2011,squire2020ocean,golden_modeling_2020}.
Major international research programmes have focussed on wave--ice interactions and the marginal ice zone where wave--ice interactions occur \citep{kohout2014storm,kohout2020observations,thomson_overview_2018,vichi2019effects}.
A central theme of the research advances is integration of wave--ice interactions into numerical wave models \citep{doble_wave_2013,ardhuin_wave_2018,meylan_threedimensional_2020,rogers_estimates_2021} and sea ice models \citep{bennetts2017brief,roach_advances_2019,bateson_impact_2020,boutin2021wave} used for operational forecasts, climate studies and other scientific investigations.
Following the framework set by \cite{dumont_wave-based_2011} and \citet{williams_waveice_2013a,williams_waveice_2013b}, wave--ice interaction models tend to couple the two key processes of (i)~wave attenuation over distance travelled due to ice cover, and (ii)~wave-induced breakup of continuous ice covers (very large ice floes) into collections of smaller floes, notwithstanding other potentially important wave--ice interaction processes, such as wave-induced ice drift \citep{grotmaack_wave_2006,williams2017wave}, wave-driven floe--floe collisions \citep{yiew_wave-induced_2017,herman_waveinduced_2018} and wave overwash of floes \citep{massom_antarctic_2010,skene_bennetts_meylan_toffoli_2015}.
Rapid progress has been made in understanding and modelling wave--ice interactions, but conspicuous knowledge gaps still exist, particularly for wave-induced ice breakup. 

Wave attenuation in the ice-covered ocean is measured using wave-buoy arrays \citep{montiel_attenuation_2018,kohout2020observations}, synthetic aperture radar \citep{ardhuin_estimates_2015,sutherland_airborne_2018} and stereo-imaging techniques \citep{campbell_observations_2014,alberello_extreme_2021}.
Analysis of the data collected suggests each frequency component of the wave spectrum attenuates exponentially with distance travelled, and that the rate of attenuation increases with increasing frequency, so that the spectrum skews towards low frequencies \citep{meylan2014situ,montiel_attenuation_2018}.
The observations of frequency-dependent exponential wave attenuation are consistent with prevailing theories, which are broadly divided into two categories: (a)~scattering theory, in which attenuation results from an accumulation of partial wave reflections by each individual floe; and (b)~viscous theories, in which the ice cover is typically treated as a continuum and wave energy is dissipated by some (often unspecified) mechanism in the ice or underlying water.
Scattering theory is applicable when wavelengths are comparable to floe sizes, and predictions show pleasing agreement with historical data from the Arctic basin where the scattering regime is generally valid \citep{kohout_elastic_2008,bennetts_three-dimensional_2010,bennetts_model_2012}.
Viscous continuum models are applicable when wavelengths are much greater than floe sizes, which is the relevant regime for most recent observations, and the viscous parameters are typically chosen to give agreement with the observations \citep{mosig_comparison_2015,cheng_calibrating_2017}.
Wave measurements are usually not accompanied by detailed measurements of the ice conditions, and the relationship between wave attenuation rates and ice conditions is largely unknown. 
However, there are consistent reports of wave blocking by continuous ice cover transitioning to relatively weak attenuation following ice breakup \citep{collins2015situ,ardhuin_ice_2020}.

Reports of wave-induced ice breakup exist in the literature, based on visual observations from ships \citep{prinsenberg_observing_2011,asplin2012fracture,kohout_situ_2016,collins2015situ} or inferred from satellite imagery \citep{ardhuin_ice_2020,herman_sizes_2021}.
Local measurements of wave and ice properties during breakup events are rare, which makes it challenging to assess the relative wave and ice conditions that lead to breakup, the resulting floe size distribution and breakup effects on waves \citep{voermans_experimental_2020}. Additional complications are presented by unknowns associated to wave attenuation \citep{bennetts_idealized_2015}.
Wave-induced ice breakup theories are typically based on stresses and/or strains imposed by a passing wave exceeding some critical value(s) \citep{crocker_breakup_1989,kohout_elastic_2008,vaughan_wave_2011}, with variants that include ice fatigue \citep{langhorne_break-up_1998}, probabilistic treatment of wave amplitudes \citep{williams_waveice_2013a}, irregular wave spectra \citep{dumont_wave-based_2011,horvat_prognostic_2015}, three-dimensional effects \citep{montiel2017modelling}, breakup memory \citep{boutin_wavesea-ice_2020} and viscoelastic ice \citep{zhang_theoretical_2021}.

Laboratory-scale physical modelling is generating understanding of wave--ice interactions to complement field observations and theory.
Wave basin experiments are used to investigate wave attenuation (and related wave transmission/reflection) due to plastic or wooden floes \citep{bennetts2015idealised,bennetts2015water,toffoli2015sea,nelli2017reflection}, freshwater ice and model ice (saline or doped), with most up to 2020 listed by \citet{parra_experimental_2020} and others since then \citep{alberello2021experimental,klein_note_2021}.
Few wave-basin experiments investigate wave-induced breakup. \citet{herman2018floe} study the floe size distribution resulting from wave-induced breakup of continuous model ice, but do not include measurements of waves in the ice-covered water. 
Therefore, they do not relate the extent of breakup to wave attenuation, and, moreover, they report breakup unexpectedly starting halfway along the length of the ice cover due to spurious waves reflected by the beach. 
\citet{dolatshah2018hydroelastic} conduct the only study \citep[to our knowledge; noting it is overlooked by][]{parra_experimental_2020}, in which both attenuation and breakup are measured simultaneously, although using freshwater ice (much stiffer than model ice) and in an essentially one-dimensional setting in a narrow wave basin.
They report a transition from wave blocking by continuous ice to relatively weak attenuation by broken ice, similar to field observations \citep{collins2015situ,ardhuin_ice_2020}.
\citet{cheng_floe_2019} report breakup in their attenuation experiments using model ice, but exclude the corresponding wave measurements from their analysis.
To date, almost all physical models of wave--ice interactions use regular incident waves, thus obscuring nonlinear processes for irregular sea-states, such as ice breakup, floe--floe collisions and overwash. 
The only exception is \citet{klein_note_2021}, who use transient wave packets, in part to avoid ice breakup.

In this article, we report a laboratory experiment in which (unidirectional) irregular wave fields are incident on model (doped) ice in a large ice tank (Aalto University, Finland).
We study the transition from a continuous ice cover to a broken ice cover using a series of tests with increasingly energetic (steep) incident wave fields, which range from causing no breakup to breaking up the entire ice-cover length.
A camera mounted above the basin monitors the ice cover during the tests, and image analysis is applied to extract the ice edge, breaking front and individual floe properties.
Wave motion in the ice-covered water is inferred from an array of bottom-mounted pressure sensors and we analyse the evolution of the wave spectrum and attenuation of the spectral frequency. Correlations between wave properties and ice breakup, during its transition from compact to broken, are discussed.

%%%%%%%%%%%%%%%%%%%%%%%%
%% EXPERIMENTAL MODEL %%
%%%%%%%%%%%%%%%%%%%%%%%%

\section{Experimental model} \label{sec:wave_ice_tank_experiments}

\subsection{Facility and model ice} \label{subsec:facility}

\begin{figure}[h]
\centerline{\includegraphics[width=\textwidth]{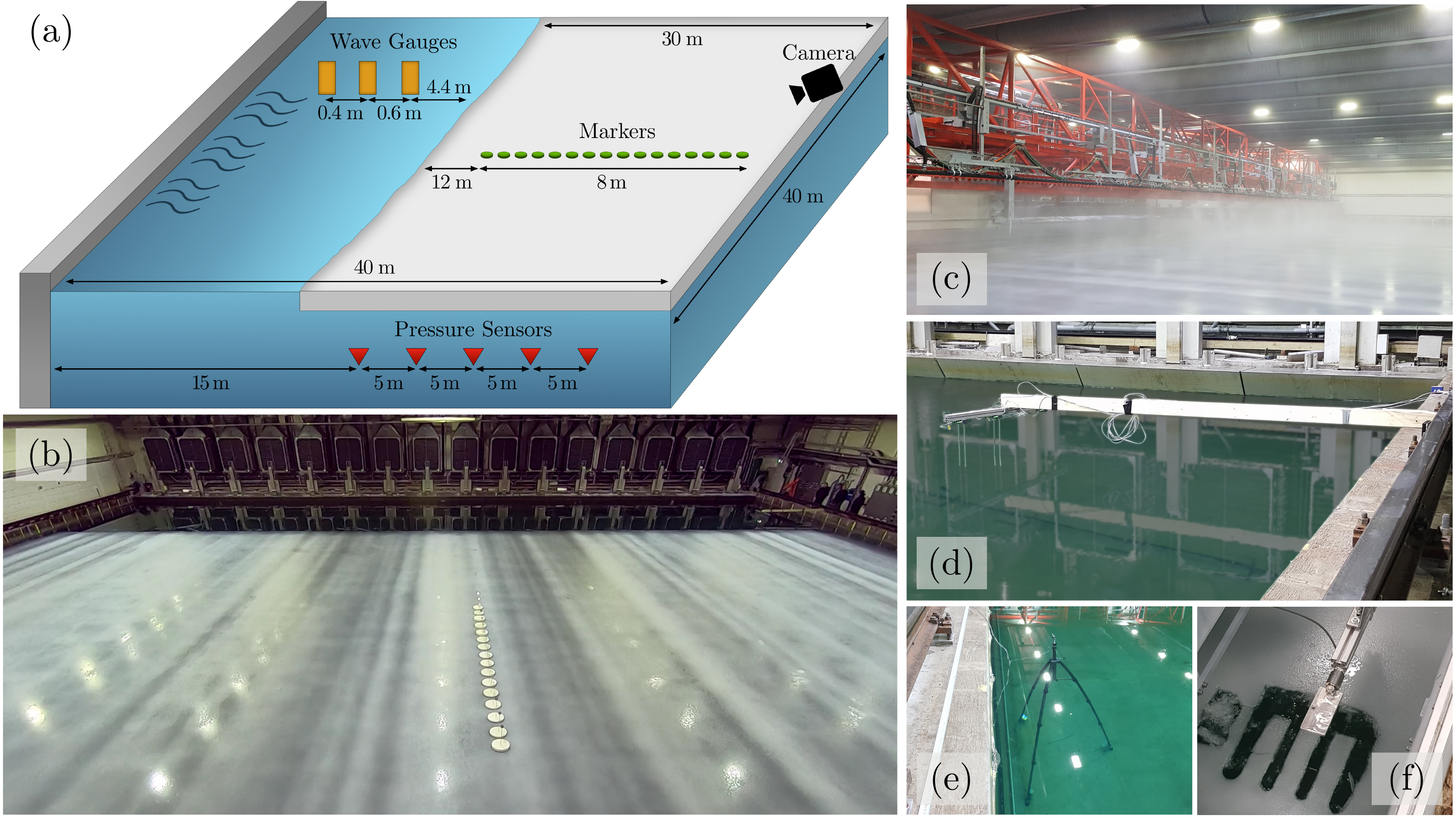}}
\caption{The physical model conducted at the Aalto ice tank: (a)~schematic of the experimental setup (not to scale) indicating the wave maker (left-hand end), adjacent open water region of length 10\,m, followed by region covered by model ice of length 30\,m (initially continuous), and instrumentation (three wave gauges in open water, five pressure sensors below ice cover, sixteen markers on the ice surface, and a video camera); (b)~image from video camera showing its field of view and the initial ice cover; (c)~{photo taken from tank side during} spray of ice crystal; (d)~photo taken from behind the wave maker showing wire wave gauges in open water; (e)~photo showing a pressure sensor before creation of model ice cover; and (f)~photo showing cantilever beam test.}
\label{fig:facility_setup}
\end{figure}

The Aalto ice tank is 40\,m long, 40\,m wide and was filled with 0.3\%-ethanol doped water up to 2.8\,m depth (Fig.\ \ref{fig:facility_setup}a). It is bounded at one end by a computer-controlled  wave-maker with 16 triangular plungers (2.5\,m wide) and by a linear beach at the opposite end to absorb incoming wave energy (95\% energy-effective for incident waves tested). The tank is equipped with a cooling system to control the ambient temperature. 

A model ice sheet with near uniform properties, and realistic thickness and flexural strength for a target scaling factor of $\approx{}30$, is produced over the entire water surface by first lowering the air temperature below freezing until a thin film of ice forms on the water surface. 
After this, layers of sub-cooled water are sprayed, forming ice crystals with diameter 1\,mm (Fig.\ \ref{fig:facility_setup}c). The ambient temperature is then lowered to $-15^{\circ}$C, which initiates crystal bonding and consolidation into an ice sheet of fine-grained structure \citep{li_preliminary_1996,von2013model}, until the desired strength is achieved. 
The ambient temperature is then increased slightly above freezing to adjust the flexural strength to its final value \citep[e.g.][]{von2019non}. The flexural strength can be tuned to a target value, but achieving a target stiffness is challenging \citep[e.g][]{schwarz1977new,von2015numerical,von2019non}. 
The mechanical properties of the ice are measured with destructive cantilever beam tests in the ice sheet (Fig.\ \ref{fig:facility_setup}f) in accordance with the International Towing Tank Conference (ITTC) guidelines.
The mechanical properties are preserved during the test by keeping the temperature close to the freezing point. The cantilever tests are carried out on a 10\,m strip next to the beach, and the strip is removed upon completion. The remaining 30\,m length of ice cover is cut free from the side walls, and  carefully (to avoid cracking and modifications in the ice sheet) pushed towards the beach to form an initial region of open water for waves to be generated without ice interference (Fig.\ \ref{fig:facility_setup}b). 

The freezing--warming cycle achieves the flexural strength $\sigma = 20.1$\,kPa, which corresponds to the realistic field-scale flexural strength $550$\,kPa at scaling factor 27.3.
The model ice thickness is $h=30$\,mm, which corresponds to a realistic field-scale thickness of $0.82$\,m.
The Young's modulus of the model ice is $E= 5$\,MPa, which corresponds to $0.14$\,GPa at field scale, i.e.\ one to two orders of magnitudes smaller than that of natural sea ice \cite[cf.][]{timco2010review}.
Model ice is typically too compliant \citep{schwarz1977new}, as it behaves nonlinearly in downward bending, exhibiting a significant plastic regime \citep{von2015numerical,von2019non}. 
Therefore, the model ice is expected to show weak elastic restoring forces when deflected by waves.

\subsection{Initial conditions and instrumentation} \label{subsec:wave_field}

\begin{table}
\caption{Pierson--Moskowitz incident wave spectra for three tests  (\ref{eq:jonswap}), plus initial and final ice conditions.}
\begin{tabular*}{\hsize}{@{\extracolsep\fill}lccccc@{}}
\hline
\hline \\
\ $H_{S{,0}}$ [m] & $T_{P{,0}}$ [s] & $\varepsilon$ & $L_{P{,0}}$ [m] & Initial ice & Final ice\\
\             &             &               &             & condition   & condition\\
\hline \\
\ 0.03 & 1.6 & 0.02 & 4 & unbroken & unbroken\\
\ 0.06 & 1.6 & 0.04 & 4 & unbroken & partially\\
\      &     &      &   &          & broken\\
\ 0.08 & 1.6 & 0.06 & 4 & partially & fully\\
\      &     &      &   & broken    & broken\\
\hline
\end{tabular*}
\label{tab:experimental_setup}
\end{table}

Starting from the continuous ice cover, the experiment consists of monitoring the propagation of unidirectional, irregular incident waves through the ice cover and ice breakup and drift in response to wave forcing. 
Three incident wave conditions are tested, with the test for each condition lasting 20 minutes, which is the maximum time possible before reflection contaminates measurements in the middle of the tank in ice free conditions, and, therefore, considered conservative for the tests with ice.
The input spectra are defined by the Pierson--Moskowitz fully developed sea state \citep{komen1996dynamics}
\begin{equation}
S_{0}(f)=\frac{\alpha_{J} g^{2}}{f^{5}}\exp\left[-\frac{5}{4}\left(\frac{f}{f_{P,0}}\right)^{-4}\right],
\label{eq:jonswap}
\end{equation}
where $g\approx{}9.81$\,m\,s$^{-2}$ is the constant of gravitational acceleration, $f$ the wave frequency, $f_{P,0}$ the frequency at the spectral peak and $\alpha_{J}$ is Phillip's constant. 
The peak frequency is $f_{P,0}=0.625$\,Hz in all three tests, which corresponds to a peak period $T_{P,0}=1.6$\,s and dominant wavelength $L_{P,0}=4$\,m \citep[based on the open water linear dispersion relation,][]{holthuijsen2010waves}. 
Therefore, the field-scale wavelength is approximately 110\,m. The Phillip's constant $\alpha_{J}$ was chosen to define different values of significant wave height ($H_{S{,0}} = 4\sqrt{m_{0{,0}}}$, where $m_{0{,0}}$ is the zeroth-order moment of the spectral density function), thereby defining different levels of wave steepness ($\varepsilon =  k_{P,0} H_{S{,0}} / 2$, where $k_{P,0} = 2\pi / L_{P,0}$ is the wavenumber at the spectral peak). 
The latter is a measure of the degree of nonlinearity of the wave field \citep[e.g.][]{onorato09,toffoli2010maximum} and an indicator of the strength of wave--ice interactions \citep[e.g.][]{bennetts2015idealised,toffoli2015sea}. 
The three tests use the steepness values $\varepsilon$ = 0.02, 0.04 and 0.06 in order of increasing magnitude, so that the broken ice cover increases in length over the duration of the experiment (Table \ref{tab:experimental_setup}).
Motion at the wave maker is determined by imposing complex Fourier amplitudes with moduli randomly chosen from a Rayleigh distribution around the target wave spectrum, and with phases randomly chosen from a uniform distribution over {$[0,2\pi)$} {\citep[this is a standard practice to produce random waves in the laboratory, see e.g.][]{onorato09,toffoli2015rogue}}. Benchmark tests without the ice sheet are conducted to verify the incident wave field.      

The water surface elevation along the tank is monitored with a combination of resistive wire wave gauges and pressure sensors, deployed 2.5\,m from the side wall (Fig.\ \ref{fig:facility_setup}a), recording at a sampling rate of 300\,Hz. 
Three wire gauges (Fig.~\ref{fig:facility_setup}d) are installed in the open water at distances of 4.6\,m, 5.0\,m and 5.6\,m from the wave maker (5.4\,m, 5.0\,m and 4.4\,m  from the ice edge) to monitor the incident wave field. The array configuration facilitates detection and removal of waves reflected by the ice cover \citep[see method in][]{mansard1980measurement}. 
Five pressure sensors (Fig.~\ref{fig:facility_setup}e) were mounted on tripods 0.2\,m below the water surface at 5\,m intervals, with the first probe 15\,m from the wave maker (5\,m from the ice edge).
A video camera is installed on the gantry above the beach, 5\,m from the ice surface and inclined of 45$^{\circ}$ to monitor ice conditions (Fig.\ \ref{fig:facility_setup}a).
The field of view (Fig.\ \ref{fig:facility_setup}b) captures the ice cover up to 22\,m from the ice edge.
Videos are recorded at a rate of 30 frames per second and at a resolution of $1920\times1440$ pixels. Sixteen markers are positioned on the ice cover every 0.54\,m with the first one 12\,m from the ice edge (Fig.~\ref{fig:facility_setup}b), in order to provide reference distances and allow reconstruction of dimensions during image analysis.

%%%%%%%%%%%%%%%%%
%% ICE BREAKUP %%
%%%%%%%%%%%%%%%%%

\section{Ice breakup} \label{sec:wave_induced_ice_breakup}

\subsection{Image processing} \label{subsec:image_processing} 

\begin{figure}[h]
\centerline{\includegraphics[width=\textwidth]{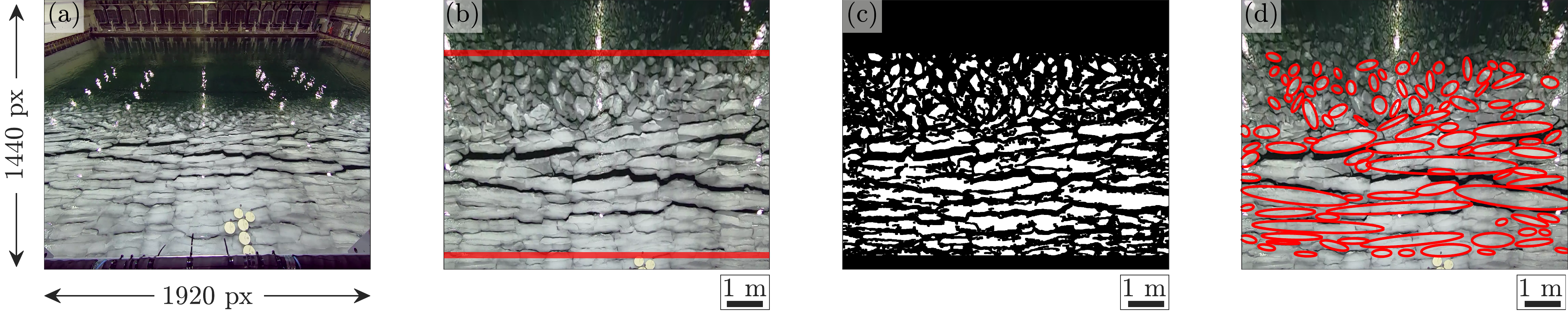}}
\caption{Stages of image processing for $\varepsilon = 0.06$ at $t \approx 300$\,s: (a)~raw frame; (b)~orthorectified frame with ice edge (upper red line) and breaking front (bottom red line) indicated; (c)~binary frame {separating ice (white) from water (black)}; and (d)~elliptical approximations for identified floes (excluding floes smaller than image resolution and large aggregate floes; red lines) superimposed on orthorectified frame.}
\label{fig:image_processing}
\end{figure}

During the tests, the continuous ice breaks up into floes and the ice edge (the boundary separating open and ice-covered water) moves.
For one video frame per second, an algorithm is applied to extract the individual ice floe geometries, the ice breaking front (the farthest line of breakup) and the ice edge. 
Image processing begins with the raw image (Fig.~\ref{fig:image_processing}a), which is orthorectified to impose a constant scale, where features are represented in their {true} positions (Fig.~\ref{fig:image_processing}b). 
The physical dimensions of the individual picture elements (pixels) are extrapolated from known distances between markers and a Gaussian filter is used to reduce white noise and improve image clarity. 
A Sobel operator is used to capture high spatial gradients of image colour \citep{jain1995machine} and identify transverse edges, which correspond to ice breakup lines parallel to the wave front and ice edge. 
Distances from the wave maker to the edges are calculated based on the number of pixels. 
An average distance is computed for each breakup line to smooth irregularities. The upper edge separating the dark water surface from the brighter ice is associated to the ice edge (top red line in Fig.~\ref{fig:image_processing}b) and the bottom edge dividing the area of broken ice from the compact sheet is identified as the breaking front (bottom red line). 

To capture individual floe geometries, the images are converted into their binary counterparts (Fig.~\ref{fig:image_processing}c), thus separating the ice from water more sharply than using the Sobel operator. 
Transverse and longitudinal edges are identified using the Canny edge detector \citep{jain1995machine}. Motivated by the elongated forms of the floes (Fig.~\ref{fig:image_processing}c), floe shapes are approximate by ellipses (Fig.~\ref{fig:image_processing}d) with the same second moment of the covariance matrix \citep{mulchrone2004fitting}. Away from the ice edge, the ellipses are oriented so that the minor axes are roughly aligned with the incident wave direction and the major axes with the wave-front direction. 

The properties of floes with areas smaller than $\approx{}0.03$\,m$^{2}$ (equivalent to a circle of diameter 0.2\,m) are uncertain due to the resolution of the images.
Therefore, these small floes are excluded from further analysis. Further, visual assessments of processed images against their raw counterparts reveals that floes with area greater than $\approx{}20$\,m$^{2}$ (equivalent to a circle of diameter 5\,m) are aggregates of smaller floes, and, hence, are not analysed. 
Overall, $\approx{}85$\% of the floes within the field of view are captured and correctly identified (e.g.\ Fig.~\ref{fig:image_processing}d).

\subsection{Ice edge and breaking front} \label{subsec:temp_evolution_ice_edge}

\begin{figure}
\centerline{\includegraphics[width=\textwidth]{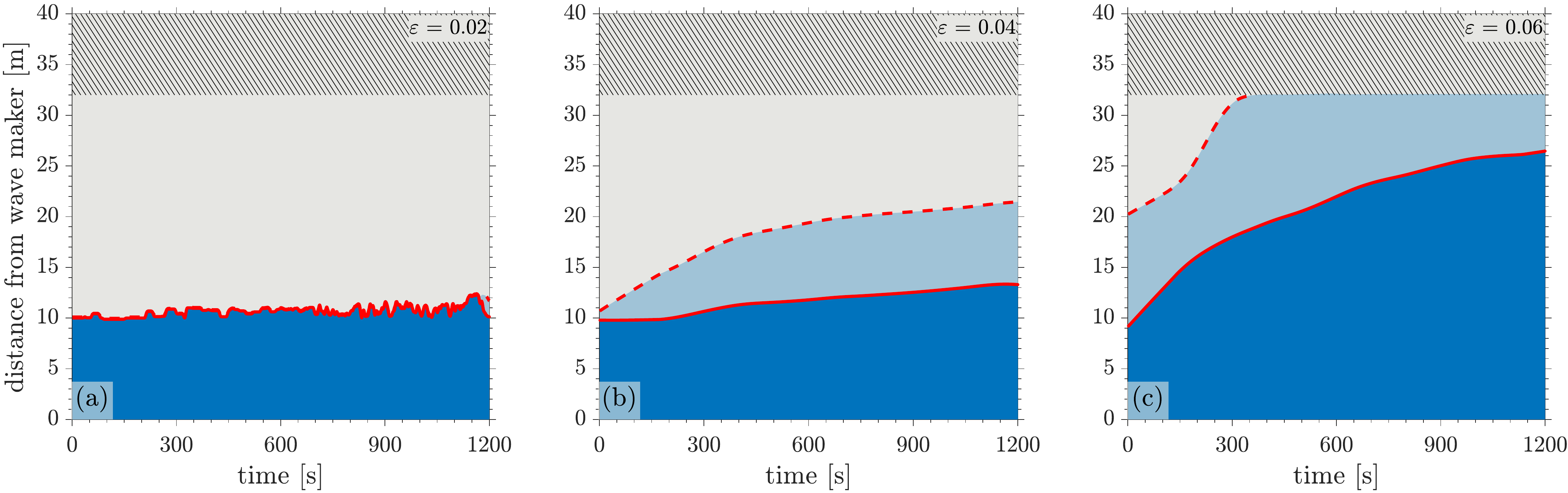}}
\caption{Spatio-temporal evolution of the ice cover during (a)~$\varepsilon=0.02$, (b)~$\varepsilon=0.04$ and (c)~$\varepsilon=0.06$ tests, indicating unbroken ice cover (white), open water (dark blue) and broken ice cover (light blue), which is bounded by the ice edge ($i_{w}$, solid red line) and the breaking front ($i_{b}$, dashed red line). The portion of the ice cover outside the camera field of view (last $\approx{}8$\,m of the tank) is hatched.}
\label{fig:ice_breaking}
\end{figure}

Fig.~\ref{fig:ice_breaking} shows the temporal evolution of the ice edge and breaking front over the three tests, denoted $i_{w}(t)$ and $i_{b}(t)$, respectively. 
The small steepness incident wave field in the first test ($\varepsilon = 0.02$; Fig.~\ref{fig:ice_breaking}a) is a lower bound for expected sea states in the Arctic Ocean \citep{de2020climate} and rarely experienced in the Southern Ocean \citep{derkani2021wind}. It propagates through the continuous ice cover without generating any visible fractures.   
The waves force overwash at the ice edge \citep{nelli2020water}, which reaches up to 1.5\,m onto the ice cover, and appears to accelerate ice edge melt and crumbling (although thermal properties were not measured), likely contributing to small variations in the location of the ice edge and breaking front (Fig.~\ref{fig:ice_breaking}a; also videos 1 and 2 in the supplementary material).
This generates the initial difference between ice edge and breaking front in the subsequent test ($\varepsilon=0.04$; Fig.~\ref{fig:ice_breaking}b).

The intermediate steepness incident field ($\varepsilon=0.04$) is common in the Arctic Ocean \citep{de2020climate} and a lower bound in the Southern Ocean \citep{derkani2021wind}. 
The ice edge breaks up as soon as the incident field reaches it, with breakup both longitudinal (parallel to the wave crests) and transverse (along the direction of wave propagation). 
The breaking front advances $\approx{}8$\,m (two wavelengths) deep into the ice cover for $t<360$\,s. The average speed of the breaking front is $\approx{}0.02$\,m\,s$^{-1}$, which is two orders of magnitude smaller than the incident group velocity ($C_{P,0} = 1.25$\,m\,s$^{-1}$, i.e.\ the velocity of the incident wave front, calculated as $0.5 L_{P,0} / T_{P,0}$ under the assumption of deep water). More generally, it means that about 62 incident wave crests penetrate into the ice in order to break one linear metre of the ice cover. For $t>360$\,s, the breaking front slows down by an order of magnitude, advancing at a speed of only $\approx{}0.003$\,m\,s$^{-1}$.  
During the second phase ($t>360$\,s), the ice edge retreats away from the wave maker due to gradual drift of the floes and ice-edge crumbling (see videos 3 and 4 in the supplementary material), and at approximately the same speed as the breaking front advances, so that the broken floe field length is almost constant, $i_{b}-i_{w}\approx{}8$\,m. 
Once the wave maker stops at the end of the test, the broken floes spread back upstream, slightly beyond the location of the initial ice edge, as indicated by {$i_{w}\approx{}9$\,m} at $t=0$ for the final test ($\varepsilon=0.06$; Fig.~\ref{fig:ice_breaking}c).

The steepest incident field ($\varepsilon=0.06$) is an upper bound for the Arctic Ocean \citep{de2020climate} and an average condition in the Southern Ocean \citep{derkani2021wind}.
Starting from a partially broken ice cover, the steep incident field breaks up the remaining continuous ice cover. Fig.~\ref{fig:ice_breaking}c only shows breakup to 32\,m from the wave maker, but visual observations confirm full breakup, such that the breaking front reaches the beach, $\approx 20$ m from its position at the beginning of this test, at $t\approx{}600$\,s.
The average speed of the breaking front is $\approx0.03$\,m\,s$^{-1}$, which is a factor 1.5 greater than the speed of the breaking front in the intermediate steepness test for $t<360$\,s (considering that the incident group velocity is unchanged, the increase of breaking front velocity means that about 40 incident wave crests are sufficient to break a metre of ice cover). 
The steep incident field also causes strong drift of the broken floes, with the ice edge moving $>15$\,m downstream over the test (videos 5 and 6 in supplementary material). 
The rate of drift slows as the test progresses and floes accumulate towards the beach.

\subsection{Floe size and size distribution} \label{subsec:floe_size}

\begin{figure}
\centerline{\includegraphics[width=\textwidth]{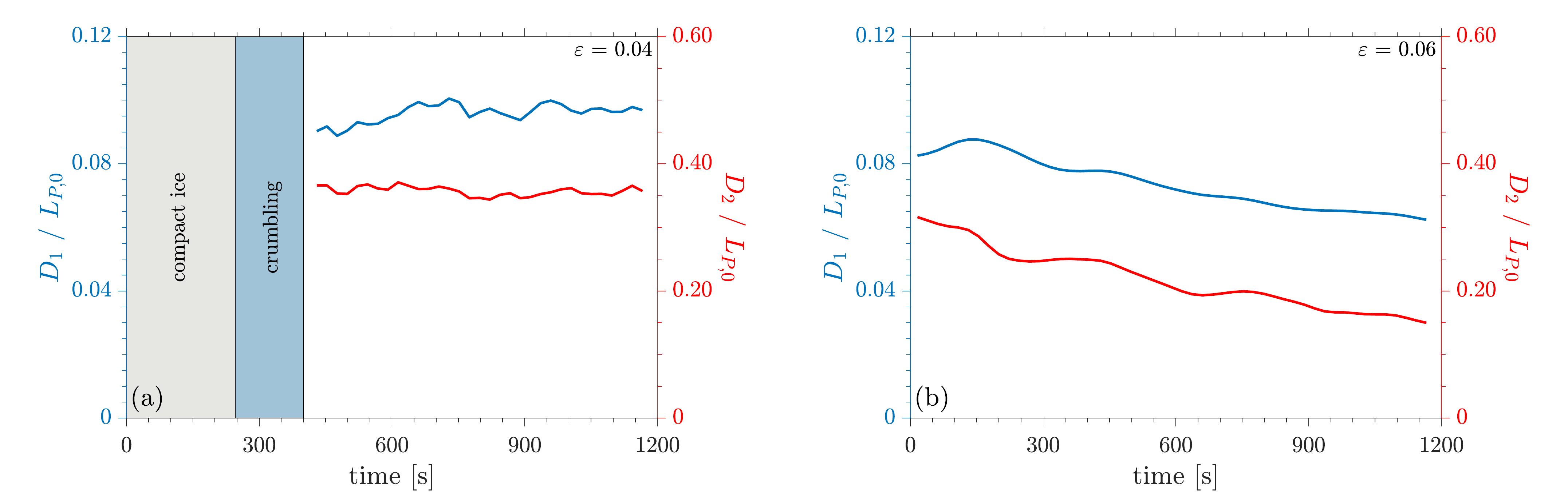}}
\caption{Temporal evolution of average minor axis ($D_{1}$; blue lines; left-hand axes) and major axis ($D_{2}$; red lines; right-hand axes) of ice floes 15--30\,m from the wave maker, and normalised with respect to the incident wavelength, $L_{P,0}=4$\,m: (a)~$\varepsilon=0.04$, (b)~$\varepsilon=0.06$ tests. Note the different ordinate axis limits used for the minor and major axes.}
\label{fig:ice_stats}
\end{figure}

Fig.~\ref{fig:ice_stats} shows evolution of the average minor axis ($D_{1}$) and major axis ($D_{2}$) of the broken floes, normalised by the dominant incident wavelength ($L_{P,0}=4$\,m), for the intermediate steepness ($\varepsilon = 0.04$; Fig.~\ref{fig:ice_stats}a) and large steepness ($\varepsilon = 0.06$; Fig.~\ref{fig:ice_stats}b) tests. 
Averages are computed using floes in the strip of the tank between 15 and 30\,m from the wave maker, thereby eliminating the ice cover beyond the field of view towards the beach and the small floes with uncertain properties close to the ice edge, for 30\,s time windows with 25\% overlaps to smooth trends. 
For the intermediate steepness, well defined floes are evident in the strip for $t>400$\,s only (videos 3 and 4 in supplementary material), and floe sizes are calculated from this point till the end of the test. 
The minor and major axes are almost constant from $t=400$\,s until the end of the test, with the minor axis $\approx 9\%$ of the incident wavelength ($D_{1}\approx{}0.36$\,m) and the major axis $35\%$ ($D_{2}\approx{}1.4$\,m). 
The large steepness incident field breaks up all of the ice in the strip, including breaking up already broken floes, causing the average minor and major axes to steadily decrease. 
At the end of the test, the minor axis is reduced to around $7$\% of the incident wavelength ($D_{1} \approx{} 0.28$\,m) and the major axis 15\% ($D_{2} \approx 0.6$\,m).

\begin{figure}[h]
\centerline{\includegraphics[width=\columnwidth]{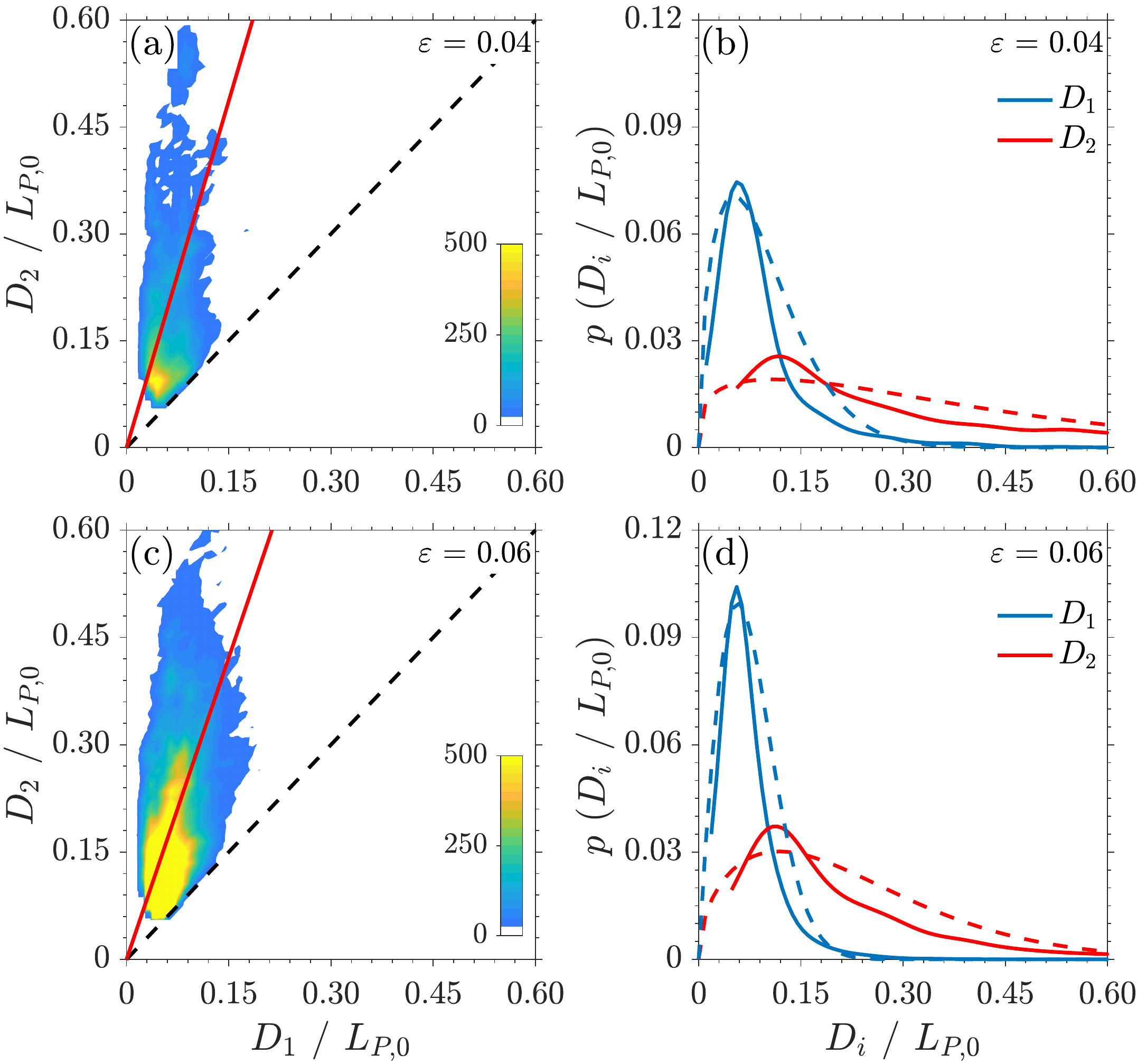}}
\caption{(a,c)~Scatter diagrams of normalised major ($D_{2}$) versus minor ($D_{1}$) axes, with linear fits (solid red lines) and 1:1 correlations (dashed black lines), for (a)~$\varepsilon=0.04$ and (c)~$\varepsilon=0.06$. (b,d)~Floe size distributions expressed as probability density functions of $D_{1}$ (solid blue lines) and $D_{2}$ (solid red lines), with corresponding fitted two-parameter Weibull distributions (dashed lines), for (b)~$\varepsilon=0.04$ and (d)~$\varepsilon=0.06$.}
\label{fig:ice_pdf}
\end{figure}

Fig.~\ref{fig:ice_pdf} shows the floe size distributions for the intermediate (top) and large steepness (bottom), as major vs.\ minor axis scatter diagrams (left panels) and probability density functions (right). The data include all floes detected from the image processing over the full 1200\,s of the tests, i.e.\ all floes detected (within the quality filter) from every image taken.
For both the intermediate and large steepness, the linear fits on the scatter diagrams reveal the aspect ratio  is approximately $D_{1}$:$D_{2} = 1$:$3$ (eccentricity $e = 2c/D_{2}\approx{}0.5$, where $c=\sqrt{(D_{2}/2)^{2}-(D_{1}/2)^{2}}$), confirming the elongated floe shapes. 
The minor axes have narrow, bell-shaped probability density functions, with a mode of 6\%, and 50\% of the floes in the 5--11\% range. 
In contrast, the probability density functions for the major axes are broad, with 50\% of floes in the range 14--47\% of the incident wavelength for $\varepsilon = 0.04$ and 11--27\% for $\varepsilon = 0.06$, and with a mode $\approx{}$12\% of $L_{P,0}$. 

The empirical probability density functions of floe sizes are linked with wave statistics via the Weibull distribution
\begin{equation}
f(D_{j}) = \frac{\beta}{D_{j}} \left(\frac{D_{j}}{\gamma}\right)^\beta \: \textrm{e}^{-\left(\frac{D_{j}}{\gamma}\right)^\beta},
\quad j=1,2,
\end{equation}
where $\gamma$ is the scale and $\beta$ is the shape parameter.
The Weibull distribution degenerates to a Rayleigh distribution for $\beta = 2$ \citep{hennessey1978comparison}, which is the statistical distribution of the incident wave amplitudes (applied at the wave maker) and, to some extent, the incident wavelengths \citep[see discussion on distribution of wave periods and wavelengths in e.g.][]{ochi2005ocean}.
The Weibull function is fitted to the empirical distributions using values of $\beta$ and $\gamma$ derived from the maximum likelihood method \citep[e.g.][]{reeve2018coastal}.
For the minor axes, $\beta = 1.6$ for $\varepsilon = 0.04$ and $\beta = 1.9$ for $\varepsilon = 0.06$, indicating the floe size in the wave direction tends to the same statistical distribution as the incident wave field as the steepness increases.
Fitting of $\beta$-values is used instead of a standard goodness-of-fit test to the Rayleigh distribution, which would only confirm fitting or rejecting of the null hypothesis.
For the major axes, $\beta = 1.3$ for $\varepsilon = 0.04$ and $\beta = 1.5$ for $\varepsilon = 0.06$, making the link between the floe size in the transverse direction and the wave statistics inconclusive.

%%%%%%%%%%%%%%%%%%%%
%% WAVE EVOLUTION %%
%%%%%%%%%%%%%%%%%%%%

\section{Wave evolution} \label{sec:wave_evolution}

\subsection{Waves-in-ice measurements} \label{subsec:pressure_sens}

The evolution of the wave field through model ice is measured indirectly by tracking the variation of pressure below the water surface at different distances from the wave maker. 
In open water, pressure variations are typically converted into water surface elevations by assuming a constant (atmospheric) pressure above the water and the linear (open water) dispersion relation \citep[e.g.][]{holthuijsen2010waves}.
The presence of ice at the water surface introduces uncertainties on both surface pressure and wave dispersion, which compromises reconstruction of the wave amplitude. 
To bypass the issue, the recorded pressure time series are normalised with the pressure standard deviation as recorded in  open water, rather than attempting to convert them into surface elevations. 
The approach provides time series that measure changes of the wave field in the ice-covered water relative to the incident field. Links between wave properties and ice breakup based on using the open water dispersion relation to convert the pressure measurements into surface elevations are discussed in \S{}\ref{sec:conclusions}.

\begin{figure}
\centerline{\includegraphics[width=\columnwidth]{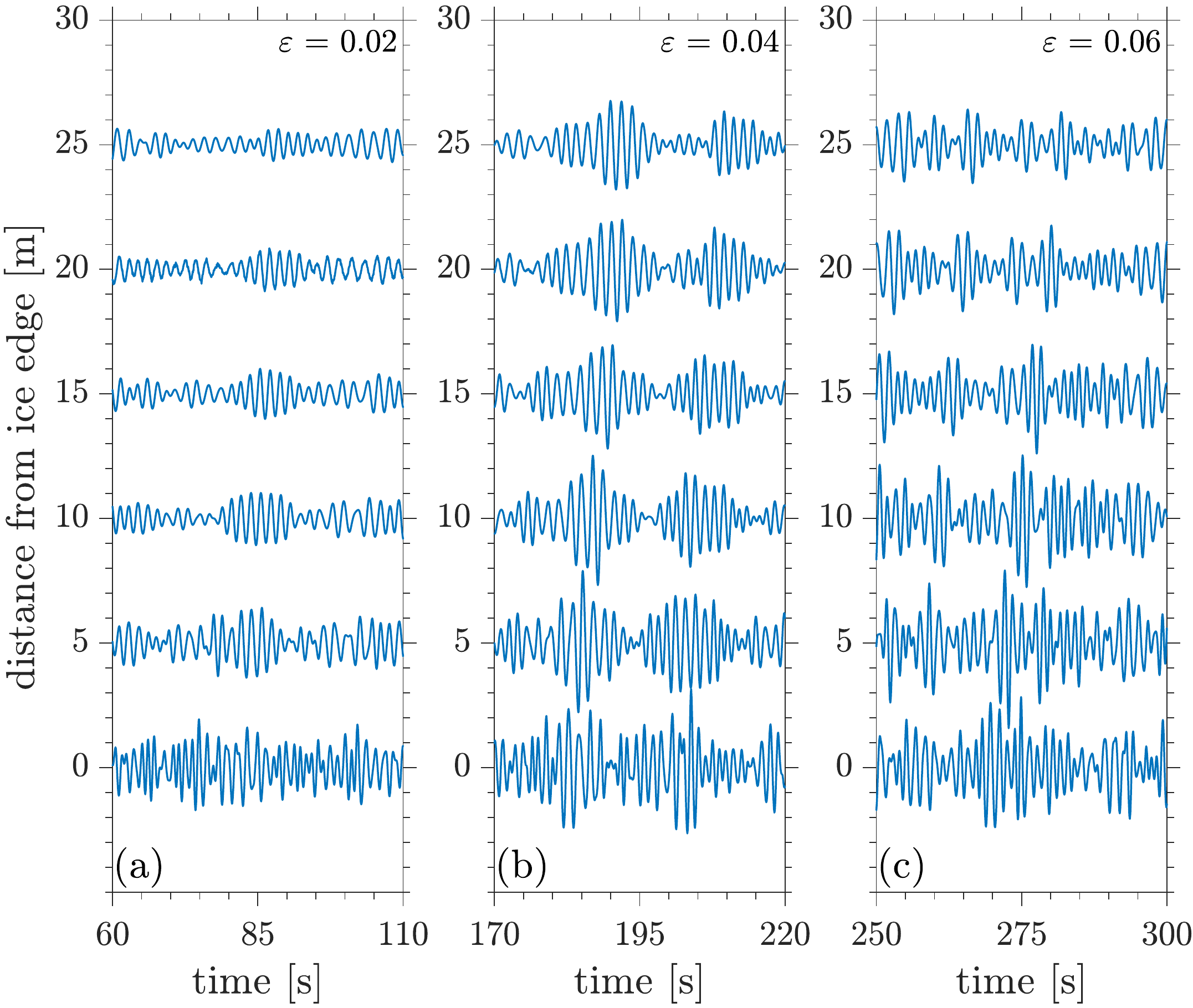}}
\caption{Time series of normalised water pressure oscillations at progressive distances from the ice edge, for (a)~$\varepsilon=0.02$, (b)~$\varepsilon=0.04$ and (c)~$\varepsilon=0.06$.}
\label{fig:elevation}
\end{figure}

Fig.~\ref{fig:elevation} shows examples of normalised pressure time series at the five different pressure sensor measurement locations and for each of the three tests conducted. 
The evolution of the series with distance into the ice-covered water are characterised by mean amplitudes decreasing, which implies reduced energy content, mean wave periods increasing, which indicates longer wavelengths, and groupiness increasing, which implies spectral narrowing \citep[e.g.][]{holthuijsen2010waves}. 
Moreover, differences are evident in the evolution of the three incident fields. The evolution of the different incident fields are quantified and compared in \S{}\ref{sec:wave_evolution}\ref{subsec:surf_elevation}--\ref{subsec:attenuation_rate}.

\begin{figure}
\centerline{\includegraphics[width=\columnwidth]{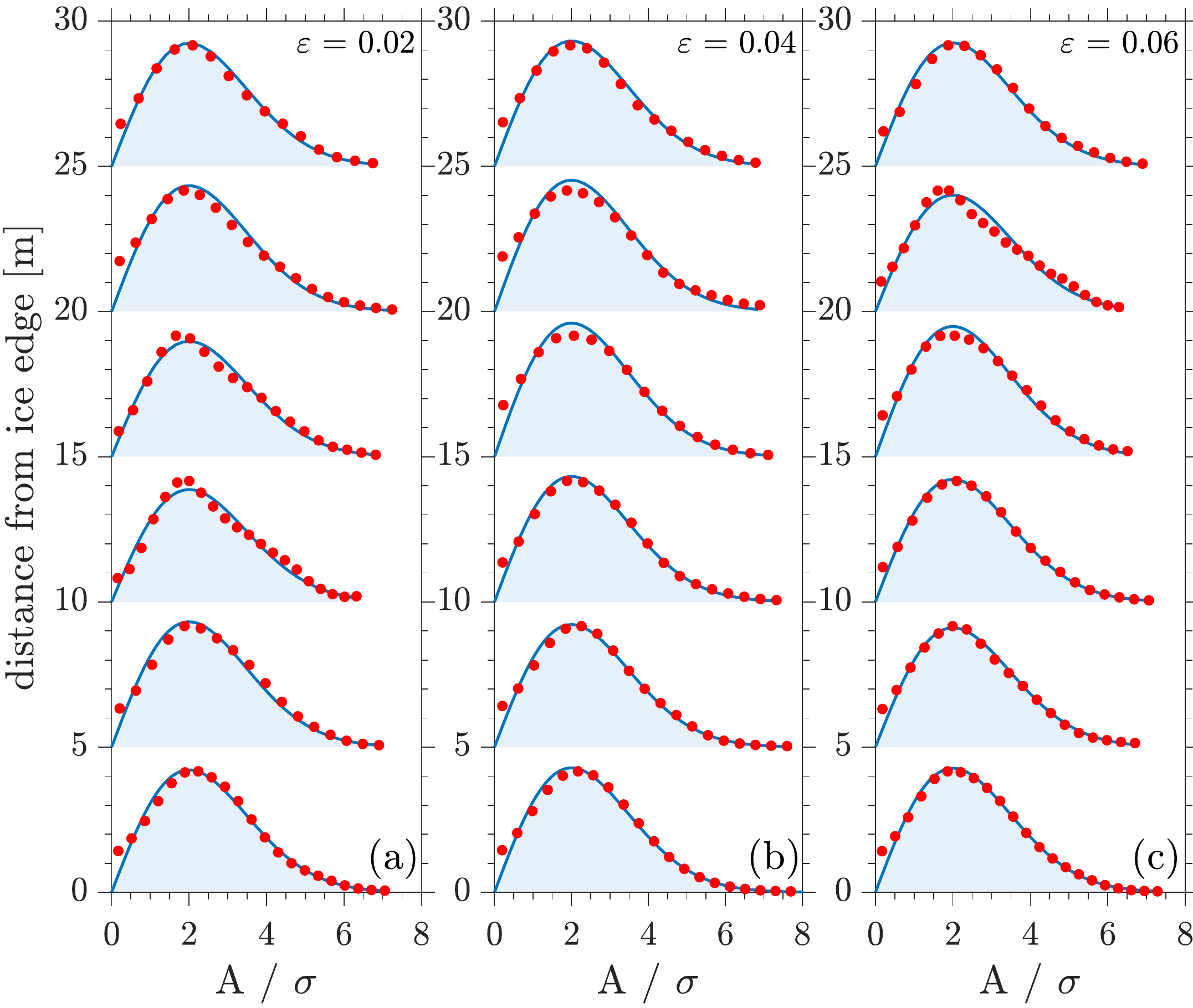}}
\caption{Empirical probability density function of the crest-to-trough amplitude ($A$) of individual oscillations of the pressure signal at different measurement locations normalised by the concurrent standard deviation $\sigma$ (full circles), and the Rayleigh distribution (blue solid line). The distributions in the bottom row are for the incident wave field in open water.}
\label{fig:waveheight}
\end{figure}

The probability density functions of the crest-to-trough amplitude ($A$) of individual oscillations of the pressure signal (normalised by the concurrent standard deviation)  are presented in Fig.\ \ref{fig:waveheight} as a function of the distance from the wave-maker, with the Rayleigh distribution included as benchmark.
The bottom panels in the figures report the statistics of the incident wave field. The amplitude distribution in open water agrees with the Rayleigh distribution, confirming it satisfies the statistics imposed at the wave maker.
Statistics remain consistent with the Rayleigh probability density function as waves propagate into the ice cover, denoting the absence of any significant nonlinear wave dynamics developing within the tank \citep[cf.][]{onorato09,toffoli2015rogue} and hence confirming linear wave physics as the relevant driver for wave evolution in the tests.

\subsection{Wave spectrum evolution} \label{subsec:surf_elevation}

\begin{figure}
\centerline{\includegraphics[width=\columnwidth]{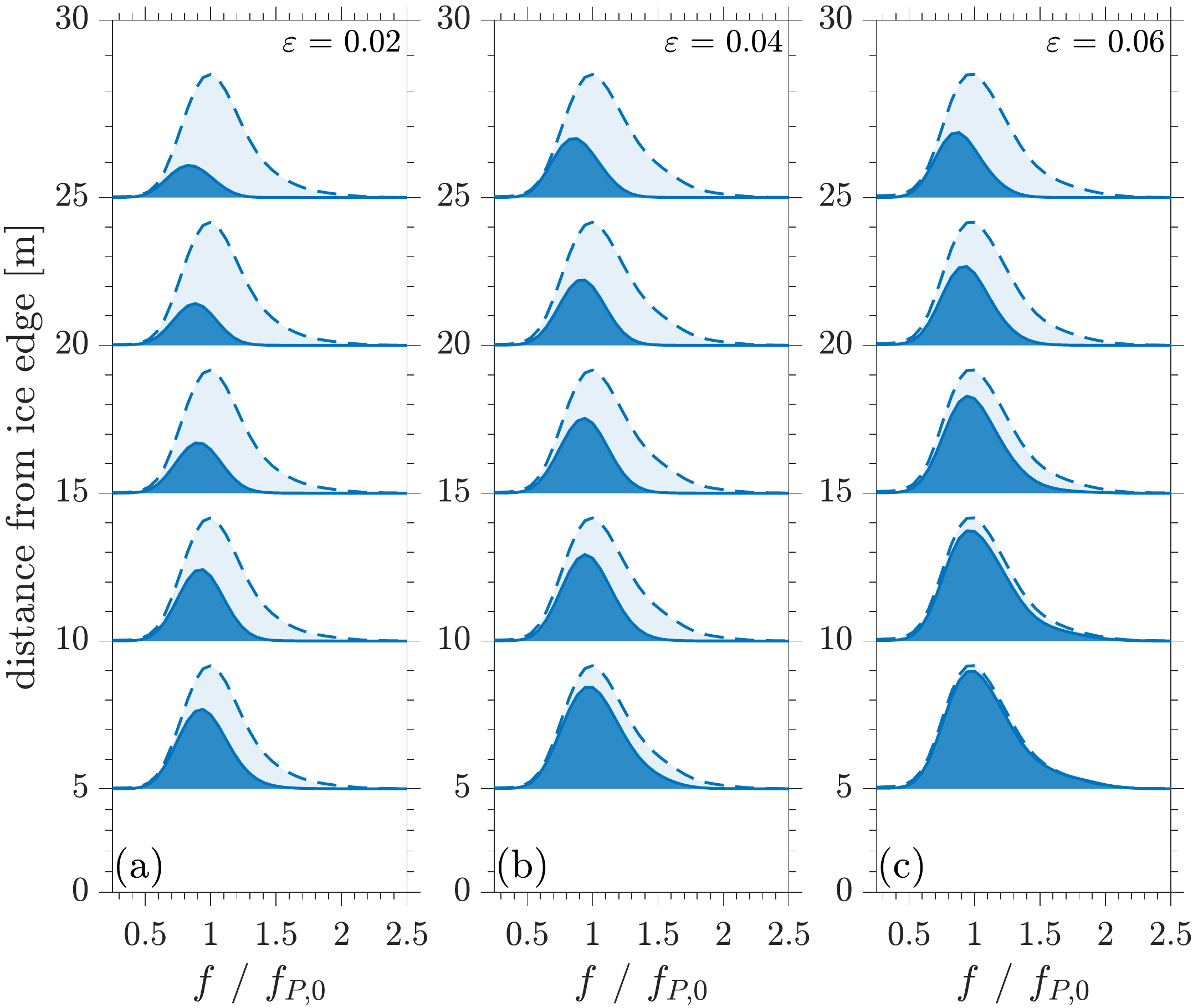}}
\caption{Wave spectra of normalised water pressure  at progressive distances from the ice edge ($S$; dark blue shade) and corresponding incident spectra ($S_{0}$; from open water benchmark tests; light blue shaded), versus normalised frequency, for (a)~$\varepsilon=0.02$, (b)~$\varepsilon=0.04$ and (c)~$\varepsilon=0.06$.}
\label{fig:wave_spectra}
\end{figure}

\begin{figure}[h]
\centerline{\includegraphics[width=\columnwidth]{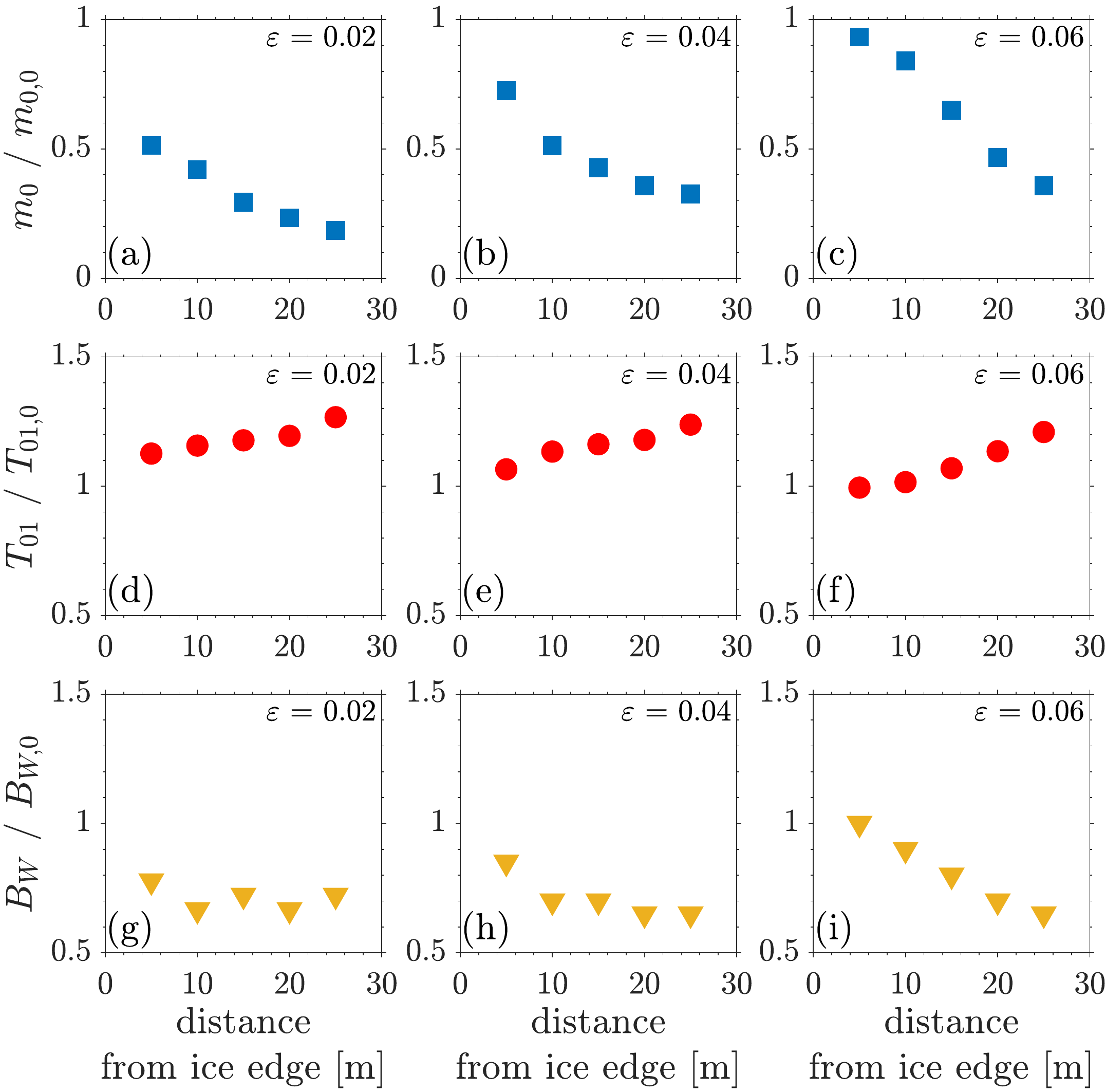}}
\caption{Wave spectrum (a--c; blue squares)~zeroth-order moments, (d--f; red bullets)~mean wave periods, and (h--i; yellow triangles)~frequency bandwidths, versus distance from ice edge, normalised by corresponding incident wave field values, for (a,d,g)~$\varepsilon=0.02$, (b,e,h)~$\varepsilon=0.04$ and (c,f,i)~$\varepsilon=0.06$.} 
\label{fig:hs_tp_bw}
\end{figure}

Fig.~\ref{fig:wave_spectra} shows the wave spectrum, $S(f)$, at each of the measurement locations and for each test, as a function of normalised frequency, $f/f_{P,0}$.
The incident spectra, $S_{0}(f)$, from the benchmark tests (without ice), are shown for reference. 
The spectra are calculated as ensemble averages of the Fourier transform of 13.7\,s intervals (4096 data points) of the normalised pressure time series with 50\% overlap.
Therefore, the spectrum at a specific location represents the change of pressure energy relative to the incident counterpart \citep[noting that the relative pressure energy is proportional to the relative wave energy, see][]{holthuijsen2010waves}.
Evolution of the spectra over distance into the ice-covered water is quantified in terms of their spectral variance ($m_0$, i.e.\ spectra integrated with respect to frequency or zeroth-order moment; due to the normalisation of the pressure time series, $m_0$ is relative to the incident counterpart and thus takes unit value in open water), the mean wave period ($T_{01} = m_0 / m_1$, where $m_1$ is the first order moment of the spectrum), and the frequency bandwidth ($B_{W}$, corresponding to the spectral width at one-tenth of energy content) in Fig.~\ref{fig:hs_tp_bw}, where each quantity is normalised by its incident wave field counterpart.

The continuous ice cover has a major effect on the energy evolution of the mild incident spectrum ($\varepsilon = 0.02$), as indicated by the corresponding time series (Fig.~\ref{fig:elevation}a).
The unbroken ice edge partially blocks the incident wave field, such that only $\approx$50\% of the incident energy remains at the first measurement location (5\,m from the ice edge; Figs.~\ref{fig:wave_spectra}a and \ref{fig:hs_tp_bw}a) and energy then reduces at a steady rate, with only $\approx$20\% of the incident energy remaining at the farthest measurement location (25\,m from the ice edge).
The sharp drop in energy at the first measurement location is attributed to reflection by the ice edge, with the reflected energy found to be $\approx$33\% of the incident energy, i.e.\ accounting for two-thirds of the energy removed by the first measurement location, and energy loss due to overwash at the ice edge \citep[see][]{nelli2020water,skene_transition-loss_2021}.
A relatively sharp energy drop at the first measurement location is also visible for the intermediate steepness ($\varepsilon = 0.04$; Figs.~\ref{fig:wave_spectra}b and \ref{fig:hs_tp_bw}b), although the drop is weaker and $\approx$70\% of the incident energy remains.
The smaller drop compared to the mild incident steepness is primarily attributed to weaker ice-edge reflection by broken ice cover ($\approx$20\% of the incident wave).
The energy drops to $\approx$50\% at the second measurement location (10\,m from the initial ice edge), which is likely due to a combination of reflection by the edge of the continuous ice cover (on the wave maker side of the sensor for $>50$\% of the test), and wave energy dissipation during ice breakup between the first and second measurement locations.
The third to fifth sensors are below continuous ice cover for the entire intermediate steepness test and the rate of energy reduction from the second to the fifth sensor is relative small, with $\approx$45\% of the incident energy remaining at the last sensor.
In contrast, for the large steepness ($\varepsilon = 0.06$; Figs.~\ref{fig:wave_spectra}c and \ref{fig:hs_tp_bw}c), almost all of the incident energy is detected at the first sensor, which is unsurprising as the ice edge retreats beyond the sensor after only $\approx{}300$\,s (the first quarter) of the test.
Similarly, over 85\% of the incident energy is recorded at the second sensor where the ice edge crosses around half way through the test.
Wave energy drops steadily and with relatively large gradient between the second and fifth measurement locations, where the waves are propagating through broken ice for the majority of the test, with $\approx$35\% of the incident energy remaining at the farthest location.

The skewing of the spectra towards lower frequencies over distance (Fig.~\ref{fig:wave_spectra}) is attributed to ice edge reflection \citep{fox_oblique_1994}, energy loss due to overwash \citep{skene_transition-loss_2021} and attenuation through continuous \citep{meylan2018dispersion} and broken \citep{bennetts_calculation_2012} ice covers being stronger for higher frequency components of the spectrum.
It follows that the mean period increases with distance into the ice-covered water (Fig.~\ref{fig:hs_tp_bw}d--f) and the spectral width decreases (Fig.~\ref{fig:hs_tp_bw}g--i). 
Similar to the wave energy (zeroth-order moment), there are large differences between the mean period and spectral bandwidth at the first measurement location and their open water counterparts for $\varepsilon=0.02$, and to a lesser extent for $\varepsilon=0.04$. 
The mean periods then steadily increase with distance. The spectral bandwidths decrease from the first to the second measurement location, and then show only a modest decrease.
For $\varepsilon=0.06$, the mean period and spectral bandwidth are indistinguishable from their open water counterparts at the first measurement location, and then increase (mean period) and decrease (spectral bandwidth) with larger gradients than in the smaller steepness tests.
For all three tests, at the farthest measurement location, the mean periods are $T_{01}=(125\%\pm{3\%})\,T_{01,0}$ and spectral bandwidths $B_{W}=(67\%\pm{4\%})\,B_{W,0}$.

\subsection{Attenuation rate} \label{subsec:attenuation_rate}

\begin{figure}
\centerline{\includegraphics[width=13cm]{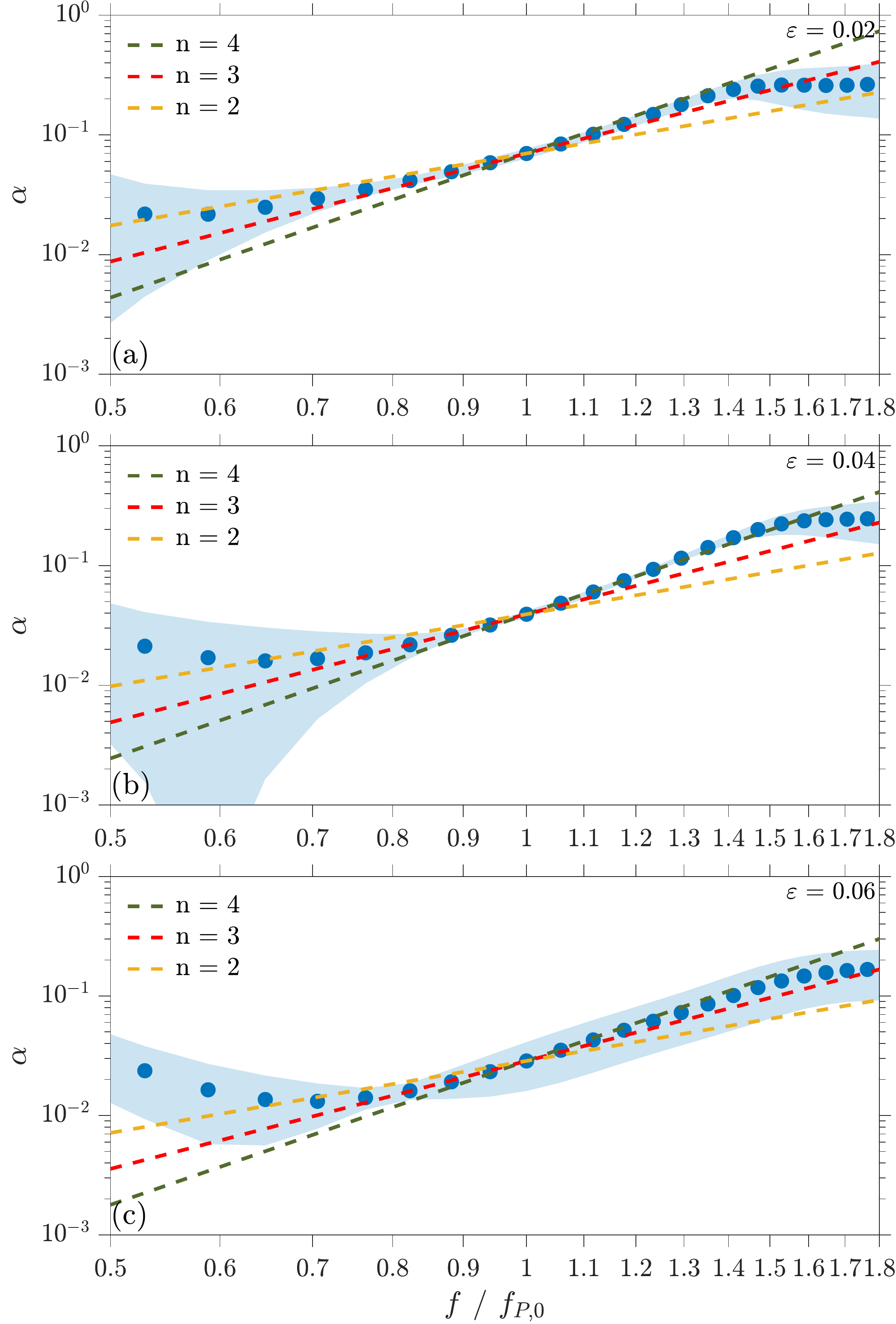}}
\caption{Empirical wave energy attenuation coefficients ($\alpha$) as functions of normalised frequency (blue circles), with 95\% confidence intervals (shaded area) and reference slopes $\alpha\propto{}f^{n}$ (dashed lines) that intersect the empirical data at $f/f_{P,0}=1$, for (a)~$\varepsilon=0.02$, (b)~$\varepsilon=0.04$ and (c)~$\varepsilon=0.06$.}
\label{fig:attenuation}
\end{figure}

Assuming each frequency component of the wave spectrum attenuates exponentially \citep{meylan2014situ} in continuous and broken ice, the attenuation coefficient (exponential attenuation rate) is
\begin{equation}
\alpha(f;d)=-\frac{\log[S(f)/S_{0}(f)]}{d},
\label{eq:attenuation}
\end{equation}
where $d$ is the distance from the initial location of the ice edge (i.e.\ 10\,m from the wave maker).
For each test, a single value of $\alpha$ per frequency is estimated using least squares regression \citep[e.g.][]{reeve2018coastal} on data from all probes. 
Fig.~\ref{fig:attenuation} shows $\alpha$ versus normalised frequency over an interval covering the most energetic part of the spectrum ($0.5 \leq f/f_{P,0} \leq 1.8$).
The 95\% confidence intervals indicate the attenuation coefficient values are, in general, robust for approximately $0.8 \leq f/f_{P,0} \leq 1.4$, over which the attenuation coefficients increase monotonically with frequency, and become uncertain towards the spectral tails, where the mean values of the attenuation coefficient are relatively insensitive to frequency. 

Field measurements (assumed to be mainly in broken on unconsolidated ice conditions) suggest a power-law dependence of the form \citep{meylan2018dispersion} 
\begin{equation}\label{eq:powerlaw}
\alpha(f)\propto{}f^{n}, 
\quad\textnormal{where}\quad 
1.9\leq n\leq 3.6,
\end{equation}
although theoretical scattering models suggest larger exponents \citep[$n\geq 8$;][]{meylan_floe_2021}.
Based on (\ref{eq:powerlaw}), reference slopes $\alpha(f)\propto{}f^{2}$ (yellow dashed lines), $\alpha(f)\propto{}f^{3}$ (red) and $\alpha(f)\propto{}f^{4}$ (green) are included in Fig.~\ref{fig:attenuation}, and set to coincide with the empirical attenuation coefficients at the spectral peaks ($f/f_{P,0}=1$).
Over the interval in which the empirical measurements are robust ($0.8 \leq f/f_{P,0} \leq 1.4$), they are, in general, bounded by the reference slopes, indicating consistency between the attenuation rate in the physical model and in field measurements.
For each incident steepness, the empirical attenuation is visibly closer to references slope $\alpha(f)\propto{}f^{3}$ and $f^{4}$  than $f^{2}$.
Based on the root mean square error over $0.8 \leq f/f_{P,0} \leq 1.4$, the empirical attenuation is marginally closer to $\alpha(f)\propto{}f^{4}$ than  $f^{3}$ for the mild and large incident steepness tests (0.0117 vs.\ 0.0162 for $\varepsilon=0.02$ and 0.0046 vs.\ 0.0066 for $\varepsilon=0.06$), and significantly closer to $\alpha(f)\propto{}f^{4}$ for the intermediate incident steepness test (0.0040 vs.\ 0.0185 for $\varepsilon=0.04$). 
Therefore, the power-law dependencies of the empirical attenuation coefficients on frequency are at the top end or just above the range derived from field measurements, which is attributed to the influence of breakup in the tests (to the best of our knowledge breakup did not occur during the field measurements) or, possibly, due to the scaling of the model ice (more compliant than natural sea ice at field scale).

%%%%%%%%%%%%%%%%%%%%%%%%%%%%%%%%
%% DISCUSSION AND CONCLUSIONS %%
%%%%%%%%%%%%%%%%%%%%%%%%%%%%%%%%

\section{Discussion and conclusions} \label{sec:conclusions}

A physical model of interactions between irregular ocean waves and sea ice conducted in an ice tank using a Pierson--Moskowitz incident spectrum and doped model ice has been reported.
Three tests were conducted, starting with a continuous ice cover, and with the wave incident steepness ($\varepsilon$) increased between tests.
The impact of the incident waves on the ice cover ranged from causing no breakup for the mildest incident field to breakup of the entire length of the ice cover and retreat of the ice edge for the largest steepness.
Breakup occurred both in the longitudinal (wave propagation) and transverse directions, and the resulting floes were shown to be well approximated by ellipses of aspect ratio 1:3.
Evidence was found of the floe size distribution in the wave propagation direction tending towards the Rayleigh distribution, similar to the distribution of incident wave amplitudes.
In contrast, the floe size distribution in the transverse direction was found to be widely spread.

Wave evolution in the ice-covered water was monitored by tracking the transformation of the subsurface pressure field with an array of bottom mounted pressure sensors.
The spatial wave evolution varied considerably between tests as inferred by wave-induced pressure attenuation, mean period increase of pressure oscillations and narrowing of the spectral bandwidth.
However, statistical properties of the wave field did not change appreciably over distance into the ice-covered water, fitting a Rayleigh distribution throughout the tank, and indicating that nonlinear wave dynamics did not develop during the tests. 
The edge of the continuous ice cover was found to have a major impact on the properties of the wave-induced pressure field (amplitude, mean period and spectral bandwidth), with differences relative to the incident condition being reduced (or removed) when the ice at the edge was broken, likely due to weaker reflection of the incident wave field.
The (exponential) energy attenuation rate in the ice-covered water showed a power-law dependence on frequency, similar to that found from field measurements and with a comparable exponent.

\begin{figure}
\centerline{\includegraphics[width=\textwidth]{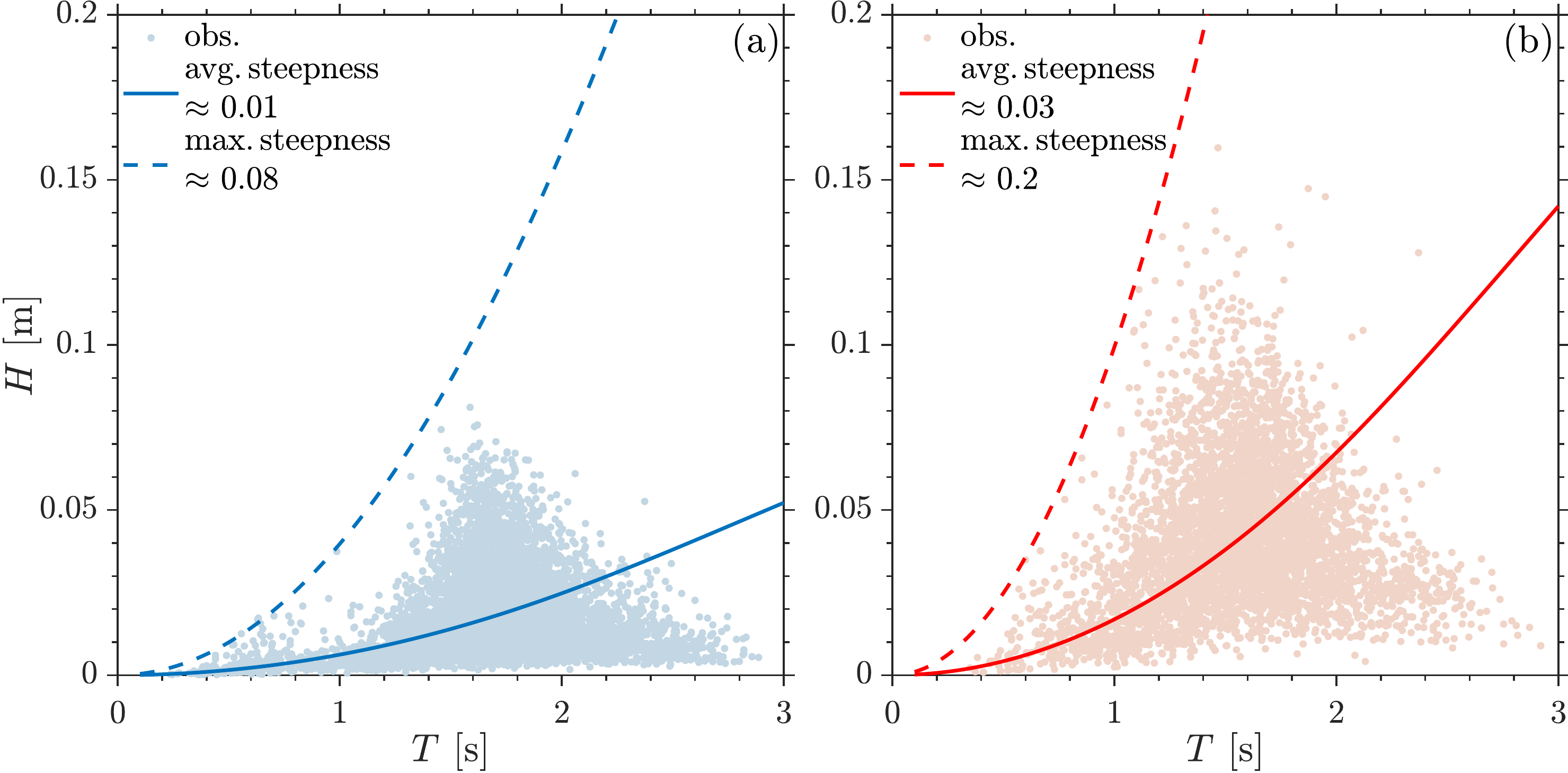}}
\caption{Empirical joint distribution of individual crest-to-trough wave heights and periods as a function of the ice configuration: (a) observations recorded in unbroken ice during the entire experimental tests (merges data obtained with incident wave fields with small and intermediate steepness, $\varepsilon$ = 0.02 and 0.04, respectively); (b) observations obtained during the transition from unbroken to broken ice (merges data obtained with incident wave fields with intermediate and high steepness, $\varepsilon$ = 0.04 and 0.06, respectively). Lines of constant steepness (average and maximum for each distribution) are reported as benchmarks.}
\label{fig:joint}
\end{figure}

Models for wave-induced ice breakup depend on the relationship between the wave height and period \citep[or wavelength; e.g.][]{dumont_wave-based_2011,williams_waveice_2013a,williams_waveice_2013b,ardhuin_wave_2018}, with breakup more likely for larger wave heights and shorter periods, i.e.\ larger steepness values.
Correlations between ice breakup and wave heights and periods in the tests are investigated by assuming the open water linear dispersion relation can be used to convert pressure into the surface elevation in the presence of ice cover \citep[see, e.g.,][for details of the conversion]{reeve2018coastal}.
A zero-crossing analysis is applied to the time series of the surface elevation to extract individual wave heights ($H$) and periods ($T$) and hence steepness, which is calculated as $kH/2$, where $k$ is the wavenumber associated to the water period ($T$). Both down- and up-crossing conditions are considered to extract a sufficiently large population \citep[see e.g.][]{reeve2018coastal}.
Fig.~\ref{fig:joint} shows the joint distribution of individual wave heights and related periods recorded at (a)~locations where the ice remained unbroken throughout a test, which includes data from the $\varepsilon = 0.02$ and 0.04 tests, and (b)~locations where ice transitioned from unbroken to broken in a test, which includes data from the $\varepsilon = 0.04$ and 0.06 tests.
Lines of constant steepness, representing the average and maximum steepness for each dataset (hence an indication of the strength of the the wave field), are included as benchmarks. For unbroken ice, wave heights do not exceed $\approx 0.07$\,m, which is about 1.5 times the significant wave height in open water, and wave periods are up to $\approx 3$\,s, which corresponds to about two times the peak period of the incident wave field. 
The wave height assumes maximum values for periods comparable with the peak period of the incident field. Wave heights tend to decrease for shorter and longer periods, with a greater decrease towards shorter periods.
Wave heights and periods distribute along a small average steepness of $\approx 0.01$, which is half the characteristic steepness of the mildest incident wave field tested (i.e.\ the joint distribution is flatter than its open waters counterpart), and are limited by an upper bound corresponding to a maximum individual wave steepness of $\approx 0.08$, i.e.\ more than five times smaller the one of a breaking wave \citep[cf.][]{toffoli2010maximum}. 
Less than 1\% of waves have steepness between 0.05 and 0.08, and, thus, wave-induced loads on the ice cover are expected to be weak, preventing breakup. 
In contrast, when the ice transitions from unbroken to broken, individual wave heights reach much larger values, with the maximum $\approx 0.15$\,m, which is about two times the incident significant wave height for $\varepsilon = 0.04$. 
The shape of the distribution more closely resembles the joint distribution in open water. The average steepness is $\approx 0.03$, which is three times as large as for the unbroken ice, and the maximum steepness is $\approx 0.2$, which is greater than the maximum for the unbroken ice by a factor of 2.5 and about half the steepness of a breaking wave. 
More than $10$\% of waves exceed the maximum steepness for the unbroken ice, indicating a more substantial load on the ice cover and justifying the transition to broken ice.

\begin{figure}
\centerline{\includegraphics[width=\columnwidth]{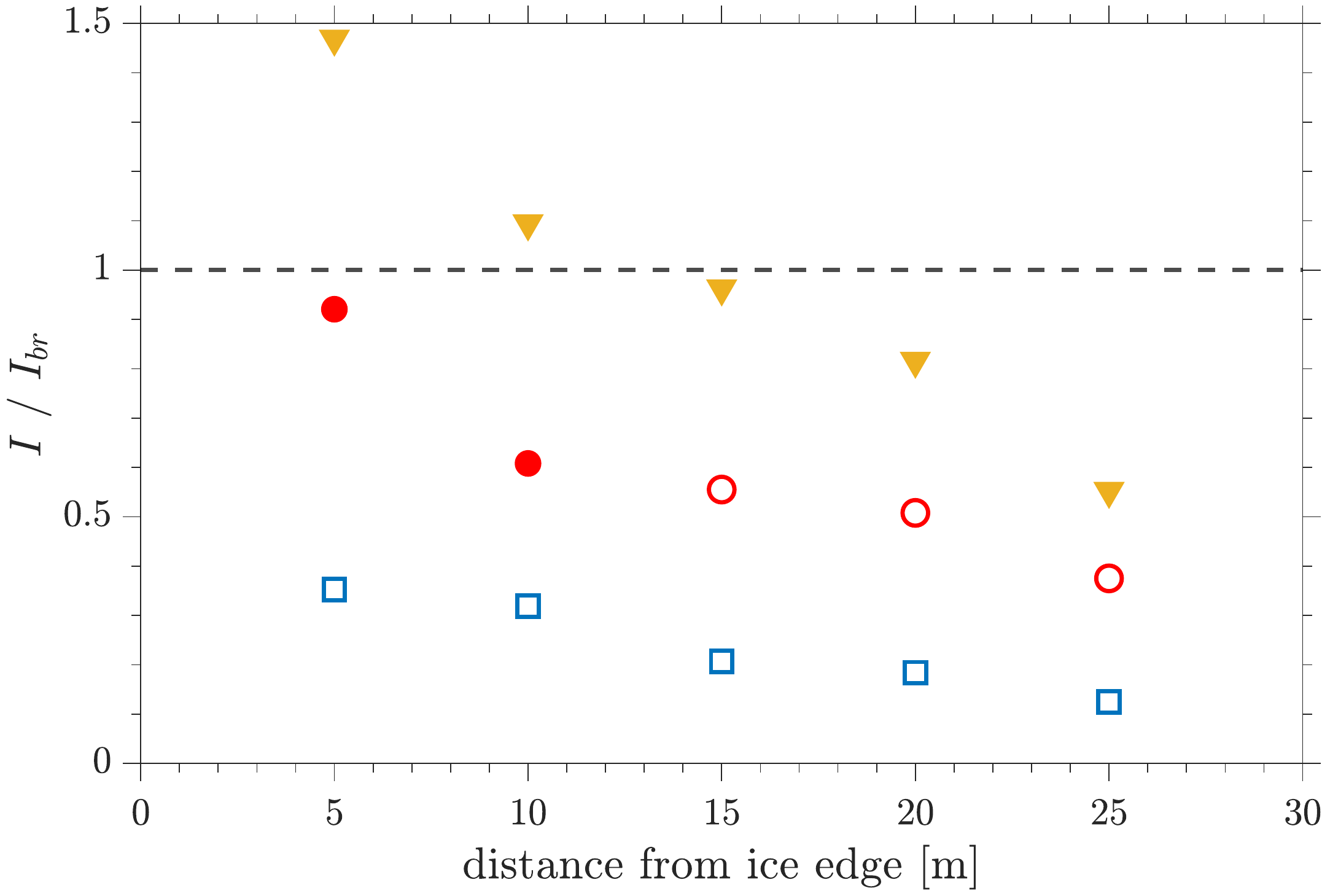}}
\caption{Breaking parameter, $I$, normalised by proposed breaking threshold $I_{br}=0.014$ \citep{voermans_experimental_2020}, versus distance from ice edge for $\varepsilon=0.02$ (blue), $\varepsilon=0.04$ (red) and $\varepsilon=0.06$ (yellow), with locations where breakup occurs (bullets) and does not occur (circles) indicated.}
\label{fig:Ibr}
\end{figure}

Results in Fig.~\ref{fig:joint} only consider the state of the ice cover, and do not incorporate information on its mechanical properties. A wave-induced breakup parameter that merges both wave and ice characteristics has been proposed as \citep{voermans_experimental_2020}
\begin{equation}
I = \frac{H_{S}\,h\,E}{2\,\sigma\,L_{P}^{2}},
\end{equation}
where the significant wave height $H_{S}$ is calculated as four times the standard deviation of the surface elevation time series at measurement locations, and the wavelength, $L_{P}$, is calculated using the open water dispersion relation at the peak period \citep[consistent with][]{voermans_experimental_2020}. Note that $I$ depends on the ratio of $H_s$ to $L_P$, which is proportional  to the wave steepness (by a factor $2\,\pi$).
Fig.~\ref{fig:Ibr} shows the values of $I$ for each test and each measurement location, normalised by the universal breaking threshold $I_{br}=0.014$ proposed by \citet{voermans_experimental_2020}, and indicating whether the ice is broken for the test and location.
In each test, the values of $I$ decrease with distance, as the significant wave height decreases and the wavelength (peak period) increases with distance.
In the small steepness test, the ice is unbroken at all locations and $I<0.55\,I_{br}$. The intermediate steepness wave field breaks the ice at the first two measurement locations, but the breaking parameter is less than the proposed threshold with $0.55\,I_{br} < I < 0.95\,I_{br}$ \citep[low values of the breaking parameter are also reported in][for experiments with model ice of different characteristics]{w13233397}.
The largest steepness breaks the ice cover at all locations and the breakup parameter spans the range $0.55\,I_{br}<I<1.47\,I_{br}$, i.e.\ it takes values less than and greater than the proposed threshold.
Note that some test and location combinations at which broken ice is indicated, the ice cover is unbroken for part of the test, and the measurements during the unbroken phases are included in the calculations of the significant wave heights and wavelengths, which adds some uncertainty to the relationship between the breakup parameter values. 
Therefore, although the value $I=0.552\,I_{br}$ at the deepest measurement location for the largest steepness test where the waves break the ice, is slightly less than $I=0.555\,I_{br}$ at the middle location and intermediate steepness test where the ice was unbroken, the results do not necessarily contradict the existence of a threshold for waves to break the ice in terms of the breakup parameter (for the particular experimental conditions).

In conclusion, the results and findings advance beyond previous physical models of coupled wave attenuation due to sea ice and wave-induced breakup \citep[particularly][]{dolatshah2018hydroelastic} by using model ice, irregular unidirectional incident wave fields and the three-dimensional nature of the facility, allowing transverse breakup.
The dataset generated new insights on (i)~the spatial and temporal evolution of the ice breakup when waves are the only forcing, and (ii)~the wave evolution and attenuation through a ice cover that transitions from a continuous to broken. This will empower assessments of combined attenuation and breakup theories.

\clearpage

%%%%%%%%%%%%%%%%%%%%%
%% ACKNOWLEDGMENTS %%
%%%%%%%%%%%%%%%%%%%%%

\textit{Acknowledgments}. $\>$
The HYDRALAB+ program supported the experiments (contract no.\ 654110).
AT and GP are supported by the ACE Foundation and Ferring Pharmaceuticals.
LGB is supported by the Australian Research Council grant FT190100404.
LGB and AT are supported by the Australian Research Council grant DP200102828.
AA, LGB and AT are supported by the Australian Antarctic Science Program Project 4434.
AT, AD, FvBuP and GP are supported by the Australia--Germany Joint Research Cooperation Scheme Project 5744547.
AA was supported by the Japanese Society for the Promotion of Science (PE19055).
GP, AA, AD and AT acknowledge technical support from the Air-Sea-Ice Lab Project.

%%%%%%%%%%%%%%%%%%%%%%%%%%%%%%%%%
%% DATA AVAILABILITY STATEMENT %%
%%%%%%%%%%%%%%%%%%%%%%%%%%%%%%%%%

\textit{Data availability statement}. $\>$
The data set is available from the corresponding authors upon request.

%%%%%%%%%%%%%%%%
%% REFERENCES %%
%%%%%%%%%%%%%%%%

\bibliographystyle{unsrtnat}

%\bibliography{references.bib}

\end{document}